\documentclass[12pt,a4paper]{article}

\usepackage{amsmath,amsfonts,amssymb,bm}

\usepackage{graphicx} 

\usepackage[hang,normalsize,bf]{subfigure} 

\begin{document}

\begin{center}

{\Huge \bf A two-component model \\ for $\gamma^*$-p
scattering \\ at small Bjorken x}

\vspace {0.6cm}

{\large T. Pietrycki $^{2}$ and A. Szczurek $^{1,2}$}

\vspace {0.2cm}

$^{1}$ {\em Institute of Nuclear Physics\\
PL-31-342 Cracow, Poland\\}
$^{2}$ {\em University of Rzesz\'ow\\
PL-35-959 Rzesz\'ow, Poland\\}

\end{center}

\begin{abstract}
We extend the Golec-Biernat-W\"usthoff model for virtual photon - proton
scattering to include the resolved photon component explicitly.
The parameters of the resolved photon component are taken from
the literature, while the parameters
of the dipole-nucleon interaction are fitted to the HERA data
in a selected limited range of $x$ and $Q^2$.
A good agreement with experimental data is obtained far beyond
the region of the fit.
\end{abstract}

\section{Introduction}

The recent decade of investigating deep inelastic scattering
at very small Bjorken $x$ at HERA has provided precise data for
the $F_2$ structure function or equivalently for
$\sigma^{\gamma^* p}_{tot}$ at large center-of-mass energies.
Many phenomenological analyses have been performed in order
to fit the data. The theoretical analyses can be divided
into two general classes. One group of models tries to fit the data
using the so-called dipole representation. In this approach,
initiated by Nikolaev and Zakharov \cite{NZ90}, one fits parameters
of the dipole-nucleon interaction \cite{GBW,FKS99,KD00} as a function
of the transverse quark-antiquark distance. Another group of models
uses rather the momentum representation \cite{MRS99,GLMN99,Schildknecht}.
Still another approach \cite{IN02} tries to fit the so-called unintegrated
gluon distributions to the HERA data (see also \cite{Sz03}).

The fits in the dipole representation take into account only
a simple quark-antiquark Fock component of the photon. However,
the higher Fock components seem to be important to understand
the diffraction \cite{GBW_glue} in detail.
The importance of the higher Fock states is at present not fully
understood. The first theoretical step in going beyond
the $q \bar q$ component has been undertaken only recently \cite{Bartels}.
However, no quantitative estimates exist up to now. Only a very
schematic QCD-inspired model was considered in Ref.\cite{BZ00}. 
On the phenomenological side, the jet production in virtual-photon-proton
scattering, especially at small photon virtuality, shows clearly
the presence of the resolved photon component
(see e.g. \cite{jets_resolved_photon}) which seems impossible to
be explained with the quark-antiquark component only.
The ratio of the dijet cross section with $x_{\gamma}^{OBS} <$ 0.75
(resolved component) to that with $x_{\gamma}^{OBS} >$ 0.75
(direct component) has been found to increase as $Q^2$ decreases. The variable
$x_{\gamma}^{OBS}$ is to be interpreted as the fractional momentum
of the photon taking part in the dijet production. At large
photon virtuality the resolved photon component disappears.
The observed $Q^2$ dependence of the resolved photon component is
roughly consistent with the naive VDM form factor \cite{VDM}.
The present NLO calculations of jet and dijet production include these
phenomenological form factors when going from real to virtual photons
(see e.g. \cite{Potter99}). Such a phenomenological factor
is then prescribed to the structure of virtual photon, and more
precisely to the parton distributions in the virtual photon.
The resolved photon component seems also crucial for understanding
the world data for the $F_2^p(x,Q^2) - F_2^n(x,Q^2)$ \cite{SU_F2p_F2n}.
All these arguments put into question the simple fits to
the total photon-nucleon cross section with the colour dipole
component alone, and call for a multi-component parametrization.

In the light of the extremely successful phenomenological description
in Ref.\cite{GBW} it is interesting to see if any phenomenological
two-component model can do a better job.
It is the aim of this note to analyze phenomenologically if
such a two-component model can satisfactorily describe the HERA
data for the photon-proton total cross section.
In our exploratory analysis, the higher Fock components are
parametrized by the standard vector dominance cross section, i.e.
our model is similar in spirit to the Bade{\l}ek-Kwieci\'nski
model \cite{BK92}.

\section{Formulation of the model}

It is known that the LO total $\gamma^* N$ cross section
in the so-called dipole or mixed representation can be written in the form
\begin{equation}
\sigma_{tot}^{\gamma^* N} = \sum_q \int dz \int d^2 \rho
\; \sigma_{T,L} \; | \Psi_{\gamma^* \rightarrow q \bar q}^{T,L}(Q,z,\rho) |^2
\cdot \sigma_{(q \bar q) N}(x,\rho) \; .
\label{dipole_nucleon}
\end{equation}
In this paper we take the so-called quark-antiquark
photon wave function of perturbative form \cite{NZ90}.
As usual, in order to correct the photon wave function for large
dipole sizes (nonperturbative region) we introduce
an effective quark/antiquark mass ($m_{eff} = m_0$).

The dipole representation (\ref{dipole_nucleon}) has been used in
recent years to fit the virtual photon - nucleon total cross section
\cite{GBW,FKS99}. The best fit has been achieved in the saturation
model of Golec-Biernat-W\"usthoff \cite{GBW}.
In their approach the dipole-nucleon cross section was parametrized as
\begin{equation}
\sigma(x,\rho) = \sigma_0
\left[ 1 - \exp\left( - \frac{\rho^2}{4 R_0^2(x)} \right) \right] \; ,
\label{GBW_dipole_nucleon}
\end{equation}
where the Bjorken $x$ dependent radius $R_0$ is given by
\begin{equation}
R_0(x) = \frac{1}{1 GeV} \left( \frac{x}{x_0} \right)^{\lambda/2} 
 \; .
\label{GBW_radius}
\end{equation}
Model parameters (normalization constant $\sigma_0$
and parameters $x_0$ and $\lambda$) have been determined by the fit
to the inclusive data on $F_2$ for $x <$ 0.01 \cite{GBW}.

In the GBW approach, the dipole-nucleon cross section is parametrized
as a function of Bjorken $x$. As discussed in \cite{Sz02},
it would be useful to have rather a parametrization in the gluon
longitudinal momentum fraction $x_g \ne x$ instead of the Bjorken $x$.
Then one could use the unintegrated gluon distribution which
is related to the dipole-nucleon cross section as
\begin{eqnarray}
\sigma_{(q \bar q) N}(x_g,\rho)
&=&
\frac{4 \pi}{3} \int \frac{d^2 \kappa_t}{\kappa_t^2}
\left[1 - \exp(i\vec{\kappa_t} \vec{\rho}) \right]
\alpha_s {\cal F}(x_g,\kappa_t^2) \nonumber \\
&=& \frac{4 \pi^2}{3} \int \frac{d \kappa_t^2}{\kappa_t^2}
\left[ 1 - J_0(\kappa_t \rho) \right]
\alpha_s {\cal F}(x_g,\kappa_t^2) \; .
\label{Fourier_transform}
\end{eqnarray}
Therefore, we find it more appropriate to parametrize the dipole-nucleon cross 
section as $x_g$ instead of the Bjorken $x$.
This involves the following replacement in Eq.(\ref{GBW_dipole_nucleon})
\begin{equation}
\sigma(x,\rho) \rightarrow \sigma(x_g,\rho) \; ,
\label{x_xg}
\end{equation}
which means also a replacement of  $x$ by $x_g$ in
Eq.(\ref{GBW_radius}).
As discussed in \cite{Sz02}, an exact calculation of $x_g$ in the dipole
representation is, however, not possible, and we approximate 
$x_g \rightarrow (M_{qq}^2 + Q^2)/(W^2 + Q^2)$, where $M_{qq}^2 =
m_q^2/(z(1-z))$ with $m_q$ being effective quark mass $m_q = m_0$
for $u / {\bar u}$ and $d / {\bar d}$ (anti)quarks and $m_q = m_0 +$ 0.15
GeV for $s/{\bar s}$ (anti)quarks.
This means that in our approach the dipole-nucleon cross section
$\sigma_{(q \bar q)N} = \sigma(W,Q^2,z,\rho)$. This must be contrasted to
the approach of Ref.\cite{GBW}, where there is no $z$ dependence
of $\sigma_{(q, \bar q)N}$.

Including higher Fock components of the (virtual) photon, one could
write somewhat schematically:
\begin{equation}
\begin{split}
\sigma_{tot}^{\gamma^* N} =& \sum_q \int d \Omega_2 \;
| \Psi_{\gamma^* \rightarrow q \bar q}(\omega_2) |^2
 \cdot \sigma_{(q \bar  q)N}(\omega_2) \\
+&
\sum_q \int d \Omega_3 \;
| \Psi_{\gamma^* \rightarrow q \bar q g}(\omega_3) |^2
 \cdot \sigma_{(q \bar  q g)N}(\omega_3) \\
+& \ldots \ldots \; .
\label{Fock_deco}
\end{split}
\end{equation}
The differentials above $d \Omega_2$ and $d \Omega_3$ represent
phase space volumes for
the $q \bar q$ and $q \bar q g$ components, respectively, and
$\omega_2$, $\omega_3$ represent the corresponding sets of kinematical
variables necessary to describe the relevant process.
The second and all subsequent terms are of the type of
a resolved photon. A rigorous approach to the problem is rather
difficult \cite{Bartels} and has not been pursued numerically.

We shall not try to follow the theoretical path sketched above.
Our aim here is somewhat different. We intend to construct a simple
two-component model. One component of our phenomenological
model is the dipole $q \bar q$ component, while the other one
is meant to represent the resolved photon explicitly.
Trying to keep our model as simple as possible and inspired by
the phenomenological results mentioned in the introduction,
we wish to check if the standard vector dominance model (VDM)
contribution can be a reasonable representation of the resolved photon.
Our approach should not be understood as a replacement
of the missing terms in Eq.(\ref{Fock_deco}).
In our opinion, the VDM contribution under consideration
contains nonperturbative terms which cannot be easily generated
by the formal expansion (\ref{Fock_deco}).
However, as already mentioned in the introduction, in many exclusive
processes the VDM contribution represents phenomenologically
the resolved photon fairly well.

The cross section for the VDM component, equivalently called here
the resolved photon component, is calculated in the standard way
\begin{equation}
\sigma^{VDM}_{\gamma^* N}(W,Q^2) = \sum_V
\frac{4 \pi}{\gamma_V^2} \frac{ M_V^4 \sigma^{VN}_{tot}(W) }{(Q^2 +
  M_V^2)^2}
\cdot (1-x) \; .
\label{VDM_component}
\end{equation}
We take the simplest diagonal version of VDM with $\rho$, $\omega$ and
$\phi$ mesons included. As discussed recently in \cite{BS03}, the
contributions of higher vector states are expected to be damped.
Above the meson-nucleon resonances it is reasonable to approximate
\begin{equation}
\sigma^{\rho N}_{tot} = \sigma^{\omega N}_{tot} =
\frac{1}{2}
 \left[ \sigma^{\pi^+ p}_{tot} + \sigma^{\pi^- p}_{tot} \right] \; ,
\label{sigma_VN}
\end{equation}
with a similar expression for $\sigma_{\phi p}^{tot}$ \cite{SU00}.
A simple Regge parametrization of the experimental pion-nucleon cross
section by Donnachie and Landshoff is used \cite{DL92}.
As in \cite{SU00}, we take $\gamma$'s calculated from the leptonic decays
of vector mesons, including finite-width corrections.
The factor (1-x) is meant to extend the VDM contribution towards
larger Bjorken $x$.

\section{Fit to the HERA data}

In the previous section we presented formulae for the virtual
photon - nucleon cross section.
The relation between $\sigma_{tot}^{\gamma^* N}$ and $F_2$ is a matter
of convention. In the so-called Hand convention one obtains
\begin{equation}
\sigma_{tot}^{\gamma^* N}(W,Q^2) = \frac{4 \pi^2
  \alpha_{em}}{Q^2(1-x)}
\left( 1 + \frac{4 M_N^2 x^2}{Q^2} \right)
\cdot F_2(x,Q^2) \; .
\label{Hand}
\end{equation}
If the Gilman convention is used instead, then
\begin{equation}
\sigma_{tot}^{\gamma^* N}(W,Q^2) = \frac{4 \pi^2
  \alpha_{em}}{Q^2}
\sqrt{ 1 + \frac{4 M_N^2 x^2}{Q^2} }
\cdot F_2(x,Q^2) \; .
\label{Gilman}
\end{equation}
We transform the structure function data
from \cite{HERA_data} in the standard but approximate way
\footnote{Both prescription (\ref{Hand}) and (\ref{Gilman})
converge to the standard formula below in the limit of small $x$.}
\begin{equation}
\sigma_{tot}^{\gamma^* N}(W,Q^2) = \frac{4 \pi^2 \alpha_{em}}{Q^2}
\cdot F_2(x,Q^2)
\; .
\label{F2_to_sigma}
\end{equation}
Then we perform two independent fits to the HERA data. In fit 1, only
dipole nucleon interaction is included (see Eq.(\ref{dipole_nucleon}))
\begin{equation}
FIT1: \;\; \sigma_{tot}^{\gamma^*N} = \sigma_{dip}^{\gamma^*N} \; .
\label{fit1}
\end{equation}
In fit 2 in addition we include the resolved photon component
in the spirit of the vector meson dominance model
(see Eq.(\ref{VDM_component}))
\begin{equation}
FIT2: \;\; \sigma_{tot}^{\gamma^*N} = \sigma_{dip}^{\gamma^*N}
                               + \sigma_{VDM}^{\gamma^*N} 
 \; .
\label{fit2}
\end{equation}

In these fits we limit to 0.15 GeV$^{2}$ $ < Q^2 < $ 10 GeV$^2$.
The upper limit is dictated by the simplicity of our model.
It is known that at large photon virtualities one has to include
QCD evolution \cite{BGK02}, which is ignored in the present analysis
for simplicity.
The maximal Bjorken $x$ in the data sample included in our fit is 0.021,
and minimal W = 17.4 GeV. With the selection criterion specified, we select
159 experimental data points.

\begin{table}

\caption{Compilation of fit parameters.}

\begin{center}

\begin{tabular}{|c|c|c|c|c|c|}
\hline
 fit           & $m_0$ (GeV) & $\sigma_0$ (mb) & $x_0$ & $\lambda$ &
 $\chi^2$ \\ \hline
 FIT1          &  0.10  &   17.0   & 9.50e-4 &  0.302   & 8.125    \\
 dipole only   &  0.15  &   23.5   & 2.00e-4 &  0.268   & 4.764    \\
               &  0.20  &   36.0   & 1.95e-5 &  0.235   & 3.080    \\      
\hline
 FIT2          &  0.10  &   7.5    & 0.0238 &  0.3160  & 1.696   \\
 dipole + VDM  &  0.15  &   8.0    & 0.0194 &  0.3107  & 1.553   \\
               &  0.20  &   8.0    & 0.0198 &  0.3213  & 1.812   \\
               &  0.30  &  15.0    & 1.67e-3&  0.250   & 1.412   \\
               &  0.40  &  24.0    & 2.20e-4&  0.230   & 1.505   \\
               &  0.60  &  55.0    & 1.10e-5&  0.230   & 4.632   \\
\hline  
\end{tabular}

\end{center}

\end{table}

In Table 1 we present the model parameters obtained from the fit.
The region of small $Q^2$ is sensitive to the value of the effective
quark mass. This nonperturbative parameter is related e.g. to the
confinement and cannot be obtained from the first principles. Therefore, in
the table we show results with different values of this parameter.
From the fit we find $\sigma_0^{fit1} \gg \sigma_0^{fit2}$
for the same value of the effective quark mass.
We have extended the range of effective quark masses in fit 2
(dipole+VDM). A good quality fit can be obtained in the broad range
of $m_0$. The $\chi^2$ criterion by itself does not allow one to answer
the question which set of parameters is better. 
The value of the $\chi^2$ per degree of freedom is shown in the last
column. The value of $\chi^2$ in fit 2 (dipole+VDM) is much smaller
than that in fit 1 (dipole only), which is not acceptable
statistically. This can be taken as the evidence for resolved
photon component.

\begin{figure}[htb] 
  \subfigure[]{\label{fig_2dmapa}
    \includegraphics[width=7.0cm]{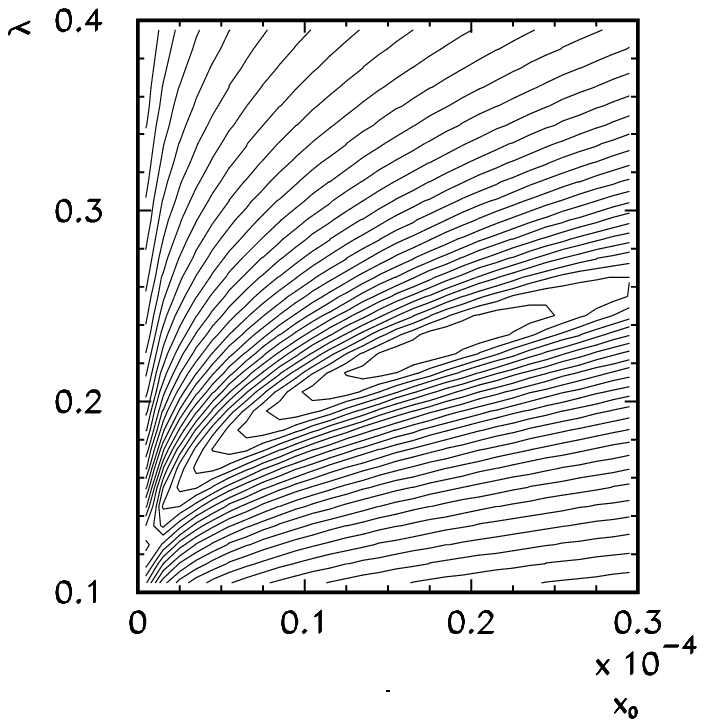}}
  \subfigure[]{\label{fig_2dmapb}
    \includegraphics[width=7.0cm]{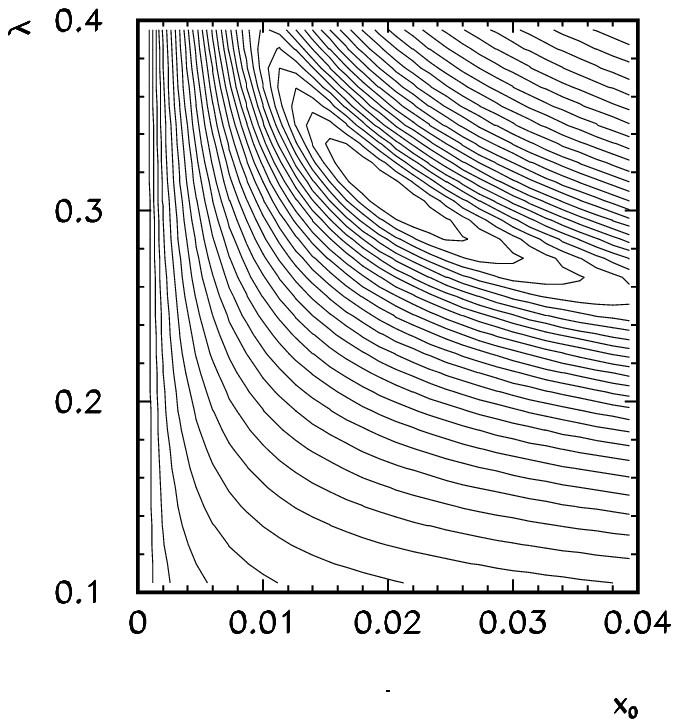}}
\caption{\it
Two-dimensional maps of $\chi^2$ per degree of freedom
for fit 1 (left panel) and fit 2 (right panel).
Please note a different range of $x_0$ for fit 1 and fit 2.
\label{fig:2dmaps}
}
\end{figure}

In order to illustrate how well the model parameters can be determined from the
fit to experimental data, in Fig.\ref{fig:2dmaps} we show two-dimensional
maps of $\chi^2$ in both cases. Here $m_0$ = 0.15 GeV and 0.20 GeV
was taken for fit1 and fit2, respectively.
Well defined minima are clearly seen. It can be seen from
Table 1 and Fig.\ref{fig:2dmaps} that the parameter $x_0$ changes
dramatically when the VDM component is included, while $\lambda$ stays
almost the same.

\begin{figure}[htb] 
\begin{center}
\includegraphics[width=8cm]{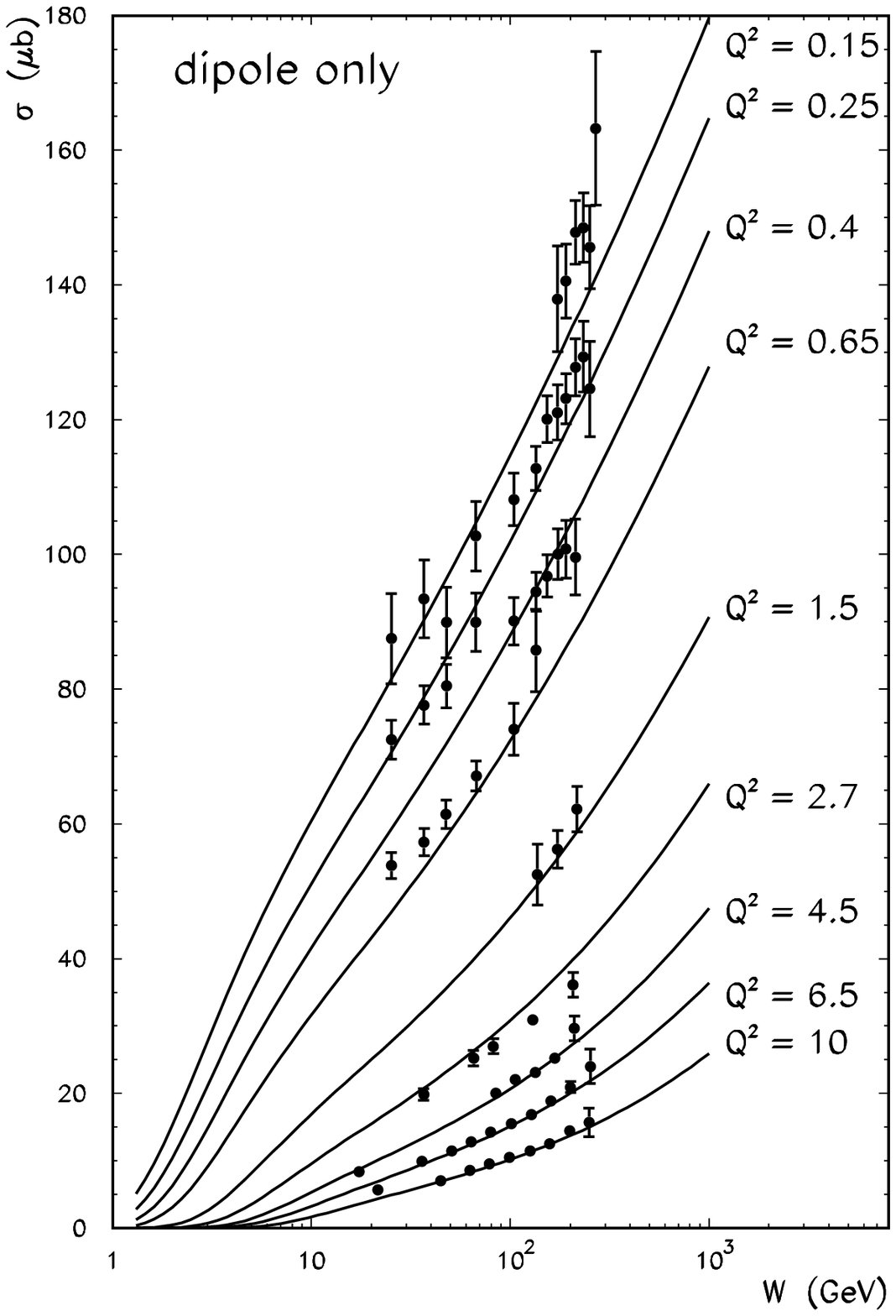}
\caption{\it
Quality of fit 1 ($q \bar q $ dipole only) -
cross sections as a function of W.
Lines and sets of experimental data are marked by the value
of photon virtuality in GeV$^2$.
The HERA data taken from \cite{HERA_data}.
\label{fig:fit1_W}
}
\end{center}
\end{figure}

\begin{figure}[htb] 
\begin{center}
\includegraphics[width=10cm]{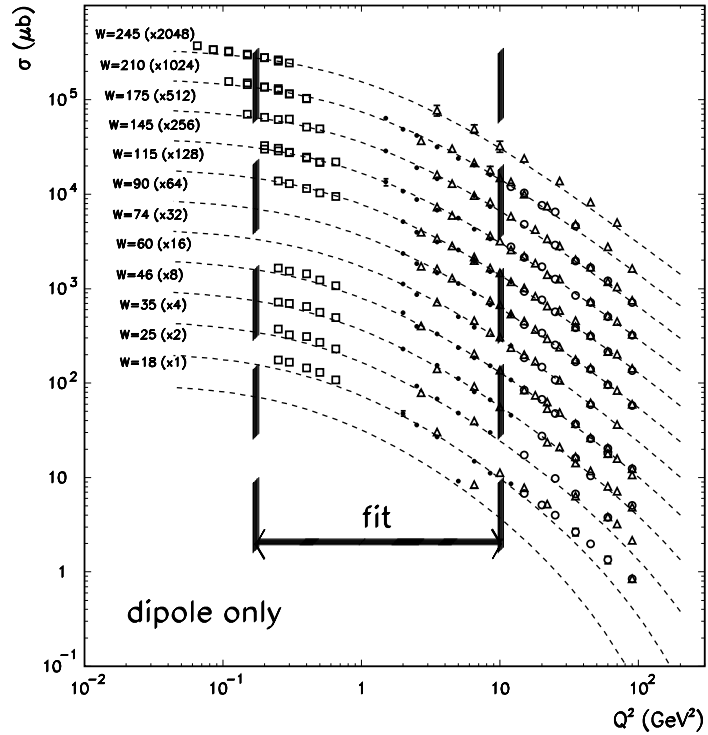}
\caption{\it
Quality of fit 1 ($q \bar q $ dipole only) -
cross sections as a function of Q$^2$.
The HERA data taken from \cite{HERA_data}.
\label{fig:fit1_Q2}
}
\end{center}
\end{figure}

\begin{figure}[htb] 
\begin{center}
\includegraphics[width=8cm]{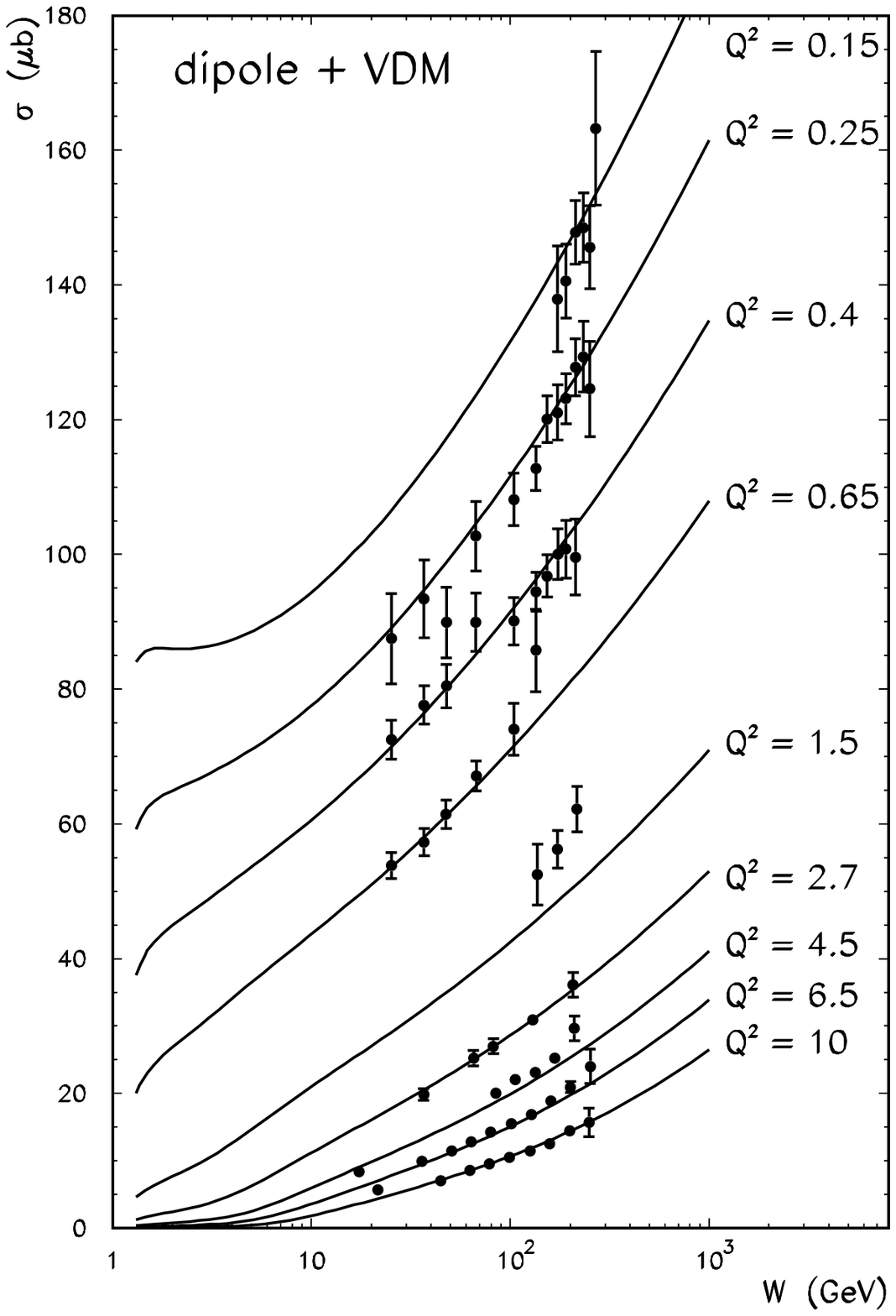}
\caption{\it
Quality of fit 2 ($q \bar q $ dipole and VDM) -
cross sections as a function of W.
Lines and sets of experimental data are marked by the value
of photon virtuality in GeV$^2$.
The HERA data taken from \cite{HERA_data}.
\label{fig:fit2_W}
}
\end{center}
\end{figure}

\begin{figure}[htb] 
\begin{center}
\includegraphics[width=10cm]{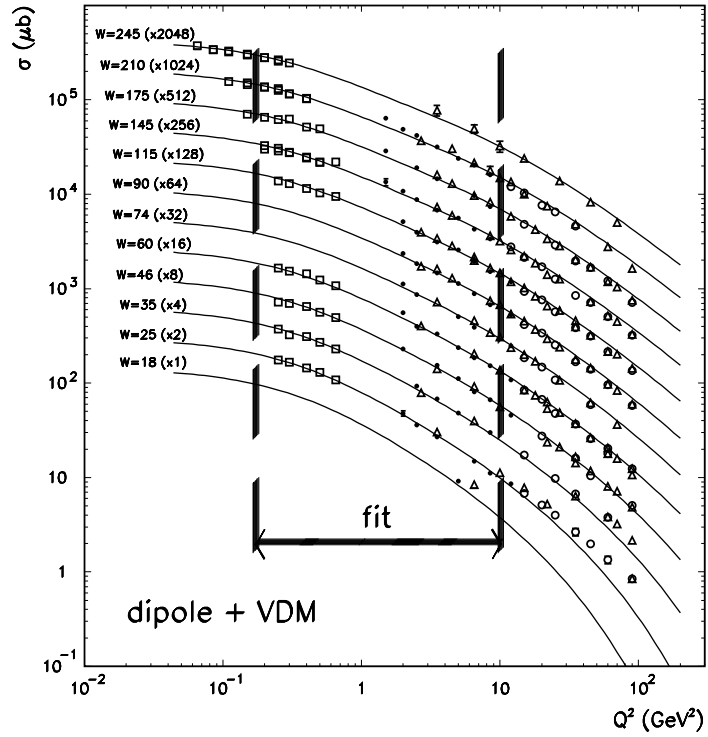}
\caption{\it
Quality of fit 2 ($q \bar q $ dipole and VDM) -
cross sections as a function of Q$^2$.
The HERA data taken from \cite{HERA_data}.
\label{fig:fit2_Q2}
}
\end{center}
\end{figure}

The quality of the fit can be judged by inspecting
Figs.\ref{fig:fit1_W} - \ref{fig:fit2_Q2}.
Since there is a rather weak dependence of the cross section on $W$,
therefore in the figures showing $Q^2$ dependence both theoretical
curves and experimental points are rescaled by an extra factor 2$^n$,
where $n$ counts the subsequent subsets of data of a given $W$
shown in Fig.\ref{fig:fit1_Q2} and \ref{fig:fit2_Q2}. Only the
cross sections for the lowest energy chosen (W = 18 GeV) are left unchanged.
By careful inspection of the figures one can see the superiority of the fit
2. In presenting the results we have made an arbitrary choice
of $m_0$. The results for other sets of parameters (different $m_0$)
are almost indistinguishable in the range of the fit. They differ
somewhat, however, outside the range of the fit where no experimental data
are available.
The theoretical curves with dipole component only underestimate
somewhat the low $Q^2$ data. We wish to stress that the quality
of our fit 1 is worse than that of the original saturation model of
Golec-Biernat-W\"usthoff \cite{GBW}. We conclude, therefore, that
parameterizing the dipole-nucleon cross section as a function of the Bjorken
$x$, instead of $x_g$, is essential for the good quality of the fit
in Ref.\cite{GBW}.

\begin{figure}[htb] 
\begin{center}
\includegraphics[width=10cm]{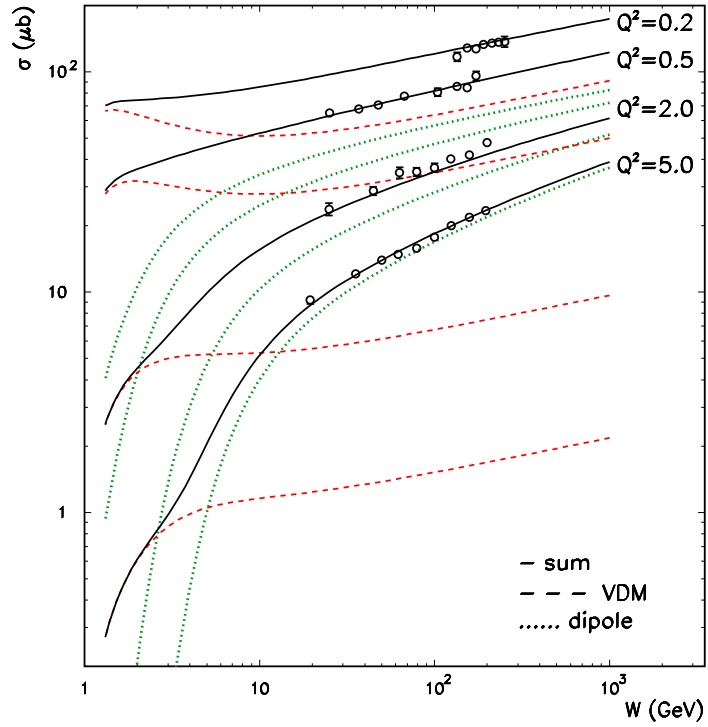}
\caption{\it
Decomposition of total $\gamma^* p$ cross section into
dipole (dotted) and VDM (dashed) contributions
for 4 different values of photon virtuality in GeV$^2$.
\label{fig:decompo1}
}
\end{center}
\end{figure}

\begin{figure}[htb] 
\begin{center}
\includegraphics[width=10cm]{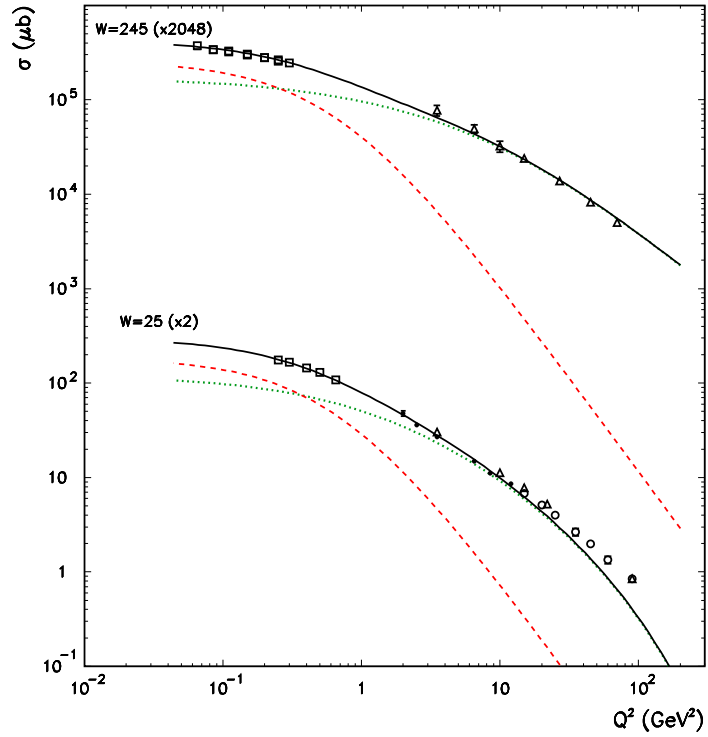}
\caption{\it
Decomposition of total $\gamma^* p$ cross section into
dipole (dotted) and VDM (dashed) contributions
for 2 different energies W.
\label{fig:decompo2}
}
\end{center}
\end{figure}

Having shown that a good-quality two-component fit to the HERA data
with very small number of parameters is possible, we wish to show 
a decomposition of the cross section into the two model components.
In Fig.\ref{fig:decompo1} and Fig.\ref{fig:decompo2} we show separate
contributions of both components as a function of $W$- and $Q^2$,
respectively. While at low energy the VDM contribution dominates
due to the subleading reggeon exchange, at higher energies they are of
comparable size. The VDM contribution, being a higher twist effect,
dominates at small values of photon virtualities.
At larger $Q^2$ the dipole component becomes dominant.
This effect is almost independent of energy.

\begin{figure}[htb] 
  \subfigure[]{\label{fig_ft1_Wa}
    \includegraphics[width=7.0cm]{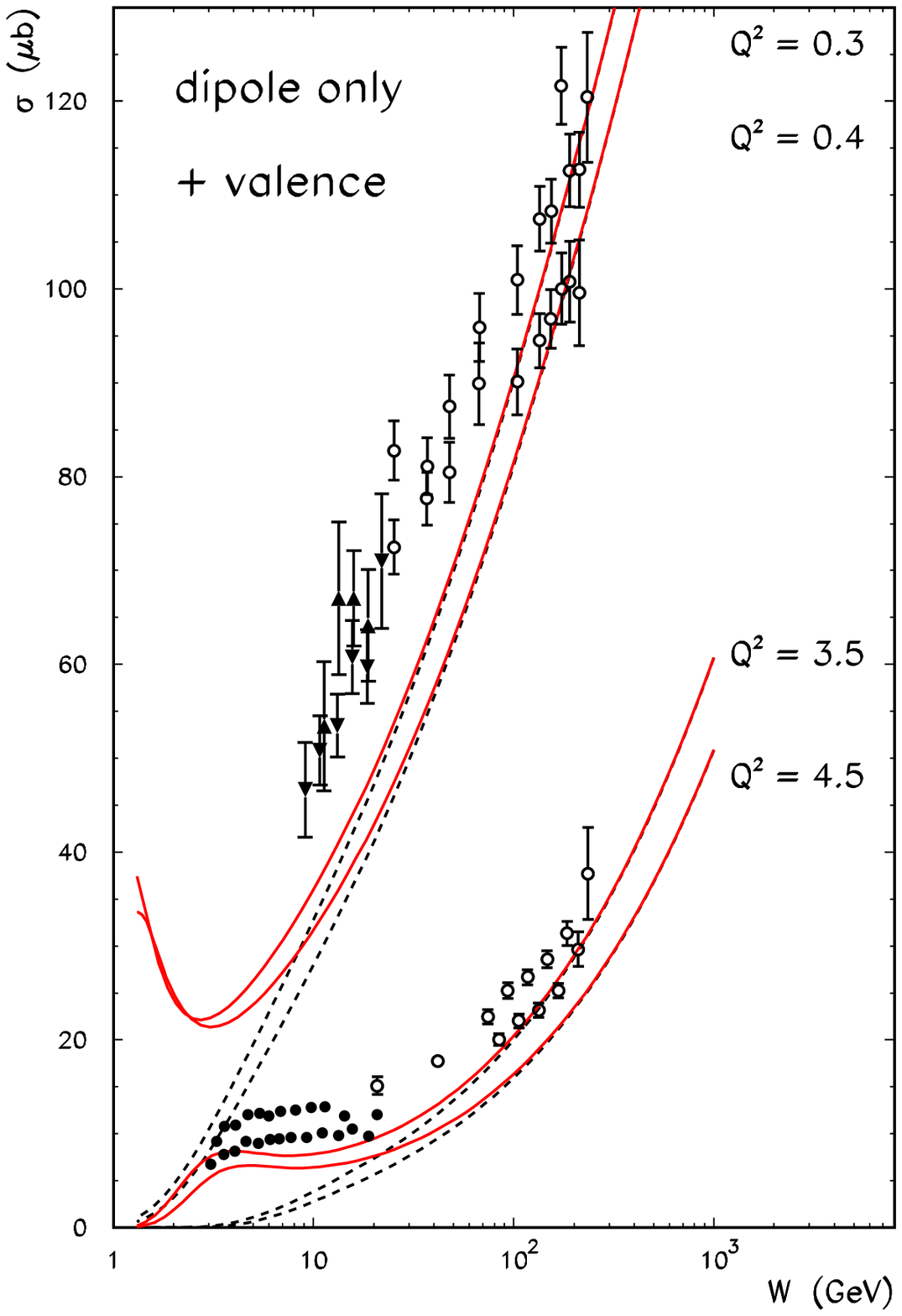}}
  \subfigure[]{\label{fig_ft2_Wb}
    \includegraphics[width=7.0cm]{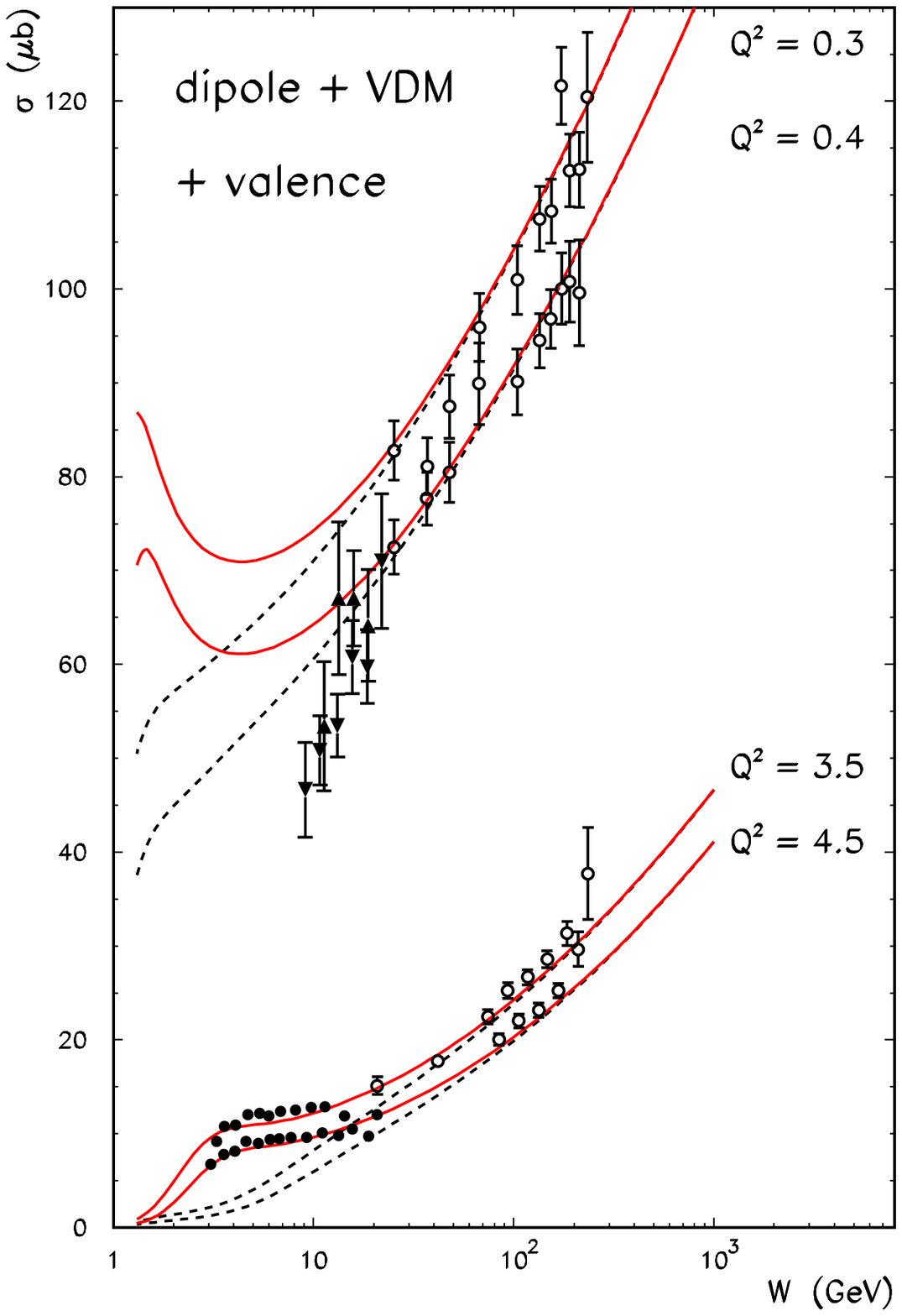}}
\caption{\it
Extrapolation of the models towards fixed target data including
valence quarks.
Solid line represent a sum of the model 1 (left panel) or model 2 (right panel)
and valence quark contribution. The dashed lines show model 1
and model 2 separately. The NMC data are shown by solid circles,
while the E665 data by solid triangles. The HERA data (open circles)
are shown for reference.
\label{fig:ft_W}
}
\end{figure}

Up to now we have concentrated on very low-x region relevant for DIS
at HERA.
It is interesting to check what happens if we go to somewhat larger Bjorken $x$
or smaller energies $W$. In this region one cannot neglect the valence
quark contribution to the cross section.
Then the cross section is a sum of three components:
\begin{equation}
\sigma_{tot}^{\gamma^* N}(W,Q^2) =
\sigma_{dip}^{\gamma^* N}(W,Q^2) +
\sigma_{VDM}^{\gamma^* N}(W,Q^2) +
\sigma_{val}^{\gamma^* N}(W,Q^2) \; ,
\label{dip_VDM_val}
\end{equation} 
where the last component is calculated according to Eq.(\ref{F2_to_sigma}) with
\begin{equation}
F_2(x,Q^2) 
\rightarrow \tilde{F}_2(x,Q^2) =
\frac{Q^2}{Q^2 + Q^2_0} 
\left ( 
\frac{4}{9} x u_{val}(x,Q^2) + \frac{1}{9} x d_{val}(x,Q^2)
\right ) \; .
\label{F2_val}
\end{equation} 
We freeze $Q^2$ below $Q^2_{min}$ = 0.25 GeV.
In the present calculation we take the leading order valence quark
distributions from Ref.\cite{GRV95}.
The $Q^2$-dependent factor in front of the r.h.s. of Eq.(\ref{F2_val})
is necessary when extrapolating the quark contribution to
the non-DIS, low-$Q^2$ region (see e.g.\cite{SU00}).
The parameter $Q^2_0$ (=0.8 GeV$^2$) is taken from a global analysis
of the experimental data in \cite{SU00}.
In Fig. \ref{fig:ft_W} we compare predictions of our two models (fits)
also with fixed target data \cite{NMC,E665}. The fixed target data
are represented by solid symbols, while the HERA data by open circles.
Formula (\ref{F2_to_sigma}) is used to calculate both experimental
and theoretical $\sigma_{tot}^{\gamma p}$ cross sections.
A better agreement is obtained with model 2 (dipole+VDM), especially at
$Q^2 \sim$ 3 - 5 GeV$^2$, i.e. for the NMC data.
An overestimation of model 2 at small energies
and small photon virtualities may be caused by neglecting
a form factor responsible for correcting the VDM contribution for
finite times of hadronic fluctuations \cite{SU00}.
Summarizing, there is a phenomenological evidence for the presence
of the resolved photon component from the analysis of experimental
data for $F_2$ in consistency with exclusive reactions.

\section{Conclusions}

Recent fits to the total $\gamma^* p$ cross section in the literature
include only the quark-antiquark component in the Fock decomposition
of the photon wave function. The contribution of
the higher Fock components, neglected so far, is not known and
difficult to calculate consistently within quantum chromodynamics.
The first trials to include the $q \bar q g$ Fock component within
perturbative QCD have not be quantified in the literature.
Nonperturbative effects, not easy to implement within the 
framework mentioned, can be also expected.
It is known from the phenomenology of the exclusive reactions that
the traditional vector dominance model in many cases gives
a good estimate of the effects characteristic for resolved photon.
In this note we have analyzed if a two-component model,
which includes the $q \bar q$ component and the more complicated
components replaced by the standard VDM, can provide
a good description of the HERA data for $\gamma^* p$ scattering.

In order to quantify the effect of the resolved photon we have performed
two different fits to the HERA data. In fit No.1 we include only the dipole
component. Here we have used the flexible and successful parametrization of 
Golec-Biernat and W\"usthoff. In comparison to their fit, in our fit
we parametrize the dipole-nucleon cross section in terms of
a variable which is closer to the gluon longitudinal momentum fraction
$x_g$ than to the Bjorken $x$.
Such a fit is useful on its own, as the corresponding unintegrated gluon
distribution can be used to estimate cross sections for many
exclusive processes. 
In fit No.2, in addition we include the VDM component
while keeping the same functional form of parametrization for
the dipole-nucleon interaction.
A better fit is obtained if the resolved photon component of the type
of VDM is included. When going to slightly larger Bjorken $x$, the model
must be supplemented for valence quark contribution. If this is done,
the model describes also the fixed target data quite well.
The two models give different predictions in the regions of
the phase space where no experimental data are available.

Our phenomenological analysis is only a first step towards a better
understanding of the role which the higher Fock components of the photon
play in both inclusive and exclusive processes.
The relation of the phenomenological VDM contribution to the formal
expansion discussed in the paper requires further study.
A numerical calculation of higher-order pQCD effects is called for
to start addressing this question quantitatively.

\vskip 1cm

{\bf Acknowledgments}
We are indebted to Jan Kwieci\'nski for a useful discussion,
and Krzysztof Golec-Biernat for a discussion
and files with experimental data.


\end{document}